\documentclass[aps,prb,twocolumn,longbibliography]{revtex4-1}

\usepackage[english]{babel}
\usepackage[utf8]{inputenc}
\usepackage{amsmath}
\usepackage{amssymb}

\usepackage{amssymb}
\usepackage{epsfig}
\usepackage{graphicx}
\usepackage{amsmath}
\usepackage{array,color}
\usepackage{natbib}
\usepackage{bm}
\usepackage{xspace}

\newcommand\shaheen[1]{\textcolor{blue}{#1}}
\newcommand{\ie}{\textit{i.e.}\xspace}
\usepackage[usenames,dvipsnames]{xcolor}
\definecolor{forestgreen}{rgb}{0.11,0.54,0.15}
\definecolor{purple}{rgb}{0.62,0.10,0.96}
\definecolor{dockerblue}{rgb}{0.11,0.56,0.98}
\definecolor{freeblue}{rgb}{0.25,0.41,0.88}
\usepackage{float}
\usepackage{breqn}

\usepackage[pdftex,plainpages=false,colorlinks=true,linkcolor=Red, citecolor=blue, urlcolor=blue]{hyperref}

\newcommand{\Tr}{\text{Tr}}

\begin{document}

\title{Interaction and temperature effects on the magneto-optical conductivity of Weyl liquids}
\author{S. Acheche, R. Nourafkan, J. Padayasi, N. Martin, and A.-M. S. Tremblay} 

\affiliation{D\'{e}partement de physique$,$ Institut quantique$,$ and
Regroupement qu\'eb\'ecois sur les mat\'eriaux de pointe$,$ Universit\'{e} de
Sherbrooke$,$ Sherbrooke$,$ Qu\'{e}bec$,$ Canada J1K 2R1}

\date{\today}
 
\begin{abstract}
 Negative magnetoresistance is one of the manifestations of the chiral anomaly
 in Weyl semimetals. The magneto-optical conductivity also shows transitions
 between Landau levels that are not spaced as in an ordinary electron gas. How
 are such topological properties modified by interactions and temperature?  We
 answer this question by studying a lattice model of Weyl semimetals with an
 on-site Hubbard interaction. Such an interacting Weyl semimetal, dubbed as Weyl
 liquid, may be realized in Mn$_3$Sn. We solve that model with single-site
 dynamical mean-field theory. We find that in a Weyl liquid, quasiparticles can
 be characterized by a quasiparticle spectral weight $Z$, although their
 lifetime increases much more rapidly as frequency approaches zero than in an
 ordinary Fermi liquid. The negative magnetoresistance still exists, even though
 the slope of the linear dependence of the DC conductivity with respect to the
 magnetic field is decreased by the interaction. At elevated temperatures, a
 Weyl liquid crosses over to bad metallic behavior where the Drude peak becomes
 flat and featureless. We comment on the effect of a Zeeman term.
\end{abstract}
\maketitle

\section{Introduction}
Weyl semimetals are three-dimensional (3D) analogs of graphene with topologically protected band
crossings and interesting transport phenomena. Monopnictides TaAs, TaP, NbAs and
NbP are prime
examples~\cite{Grassano_Pulci_Mosca:2018,WengBernevig:2015,LeeZahid:2015}. In
the presence of a uniform magnetic field, a Weyl node splits into degenerate
Landau levels with a chiral zeroth level that crosses the Fermi level and gives
a non-zero DC conductivity. The degeneracy of the zeroth Landau level depends on
the amplitude of the magnetic field. Hence, the resistivity in the presence of
parallel electric and magnetic fields acquires a magnetic-field-dependent
contribution, the so-called chiral anomaly contribution, which is negative in
contrast with the conventional metal.~\cite{Son_2013} Negative magnetoresistance
has been experimentally observed in Weyl semimetals such as
TaAs~\cite{Huang2015,Zhang_2016} or Mn$_3$Sn~\cite{Kuroda2017} even though it is
not always clear whether its origin is the chiral
anomaly.~\cite{Arnold2016,Goswami2015,Naumann_Arnold_Hassinger_2020}

The standard Drude contribution and the chiral-anomaly contribution compete with
each other in general, leading to a nonmonotonic dependence of the conductivity
(or resistivity) on magnetic field. Whether the chiral anomaly-related
conductivity dominates or not depends on the ratio of different scattering
times, such as inter-Weyl point scattering time and  transport scattering time
and the location of the chemical potential with respect to Weyl points. These
quantities are impacted by both temperature and electron-electron interaction,
in particular in correlated Weyl semimetals such as Mn$_3$Sn~\cite{Kuroda2017},
raising the question of their influence on the magnetoresistance of Weyl
semimetals.

The frequency dependence of the conductivity in Weyl semi-metals also has
interesting features. At zero temperature and in the absence of a magnetic
field, the optical conductivity in the continuum limit and without interactions
exhibits a linear frequency dependence with vanishing DC limit when the chemical
potential is at the nodes.~\cite{Ashby2013a, Ahn2017} This is a consequence of
the parabolic density of states. It has been observed in the low-temperature,
low-frequency optical spectroscopy of the known Weyl semimetal
TaAs.~\cite{Xu2016} Adding a magnetic field, not only leads to the
above-mentioned chiral-anomaly in the DC conductivity, it modifies the optical
conductivity that now consists of a narrow low-frequency peak, whose DC value
manifests the chiral nomaly, and of a series of asymmetric peaks from interband
transitions superimposed on the linear background from the no-field
case~\cite{Ashby2013a,Yuan:2018}. The case of hybridized Weyl nodes has also
been considered both theoretically and experimentally for NbP~\cite{Jiang:2018}.
The effect of long-range Coulomb interactions on interband magneto-optical
absorption has been studied theoretically using GRPA~\cite{Bertrand:2019}. 

Here, using dynamical mean-field theory~\cite{Georges_1996}, we study, based on
Ref.~\onlinecite{Acheche_PhD_thesis}, the effects of both local Hubbard-type
interactions and temperature on the magneto-optical conductivity. We show how
interactions and temperature broaden the low-frequency peak, renormalize and
redistribute optical spectral weight between the low-frequency peak and the
inter-band transitions between Landau levels and how interactions transfer
optical weight to the high-frequency incoherent satellites. Theoretically, all
this is related to the robustness of the quasiparticle picture, which we
thoroughly analyze. We will refer to the interacting Weyl semimetal as a Weyl
liquid~\cite{WeylLiquid:2018}, by analogy with the Fermi liquid.  

We also investigate the temperature-induced transition to bad-metal behavior.
This occurs as follows in normal metals. The half bandwidth of the low-frequency
peak in optical spectra, proportional to the scattering rate, grows with
increasing temperature because of thermally induced scattering events grow. Upon
approaching the so-called Mott-Ioffe-Regel~\cite{Ioffe1960}, the scattering rate
becomes of the order of the bandwidth, and the low-frequency peak becomes flat
and essentially featureless~\cite{Brown_2019, Deng_2013,
Xu_Haule_Kotliar_2013,Cha_Patel_Gull_Kim_2019}. Quasiparticles and Fermi liquid
behavior disappear. How this physics manifests itself in a Weyl liquid is
another subject of this study. 

After we introduce the model in Sec.~\ref{Sec:Model}, we show the effect of
interactions on single-particle properties in Sec.~\ref{Sec:SinglePart} and then
of interactions and temperature on the conductivity in Sec.~\ref{Sec:Cond}. Appendix~\ref{AppendixZeeman} shows that the Zeeman term has little influence on the magneto-optical conductivity, even though it shifts the Landau levels.
Appendix~\ref{Appendix1} contains a perturbative estimate of the imaginary part
of the self-energy, and shows that despite its $\omega^8$ dependence, the usual
quasiparticle renormalization $Z$ follows. Appendix~\ref{Appendix} explains how
to extract the weight of an isolated Drude peak directly from imaginary
frequency.

\section{Model and Methods}\label{Sec:Model}

We start from a Weyl semimetal model defined on a cubic lattice and add a
particle-hole symmetric Hubbard interaction. In the zero-field case, it reads
\begin{equation}\label{intHam}
\hat{H} = \hat{H}_{0} + U\sum_{\textbf{r}}(\hat{n}_{\textbf{r}\uparrow}-1/2)(\hat{n}_{\textbf{r}\downarrow}-1/2)- \mu\sum_{\textbf{r}}\hat{n}_{\textbf{r}}
\end{equation}
where $U$ is the strength of the Hubbard interaction,
$\hat{n}_{\textbf{r}\uparrow}$ ($\hat{n}_{\textbf{r}\downarrow}$) are occupation
numbers for spin up (down), and $\mu$ is the chemical potential. The
non-interacting Hamiltonian $\hat{H}_0$ in second quantized form
is~\cite{Roy2016}
\begin{align}
\label{nonintHam}
    \hat{\mathbf{H}}_0 & = \sum_{\textbf{k}}\hat{\mathbf{C}}_{\textbf{k}}^{\dagger}[-(2t\cos{k_x} + 2t\cos{ k_y} + 2t_z\cos{k_z})\sigma_{z} \nonumber\\ & + (2t\sin{k_y})\sigma_y + (2t\sin{k_x})\sigma_x]\hat{\mathbf{C}}_{\textbf{k}},
\end{align} 
where $t$ and $t_z$ are independent parameters and $\sigma_x$, $\sigma_y$ and
$\sigma_z$ represent Pauli matrices. The Hamiltonian is written in spin-space,
so the creation and annihilation operators are two-spinors,
$\hat{\mathbf{C}}_{\mathbf{k}}=(\hat{c}_{\mathbf{k},\uparrow},
\hat{c}_{\mathbf{k},\downarrow})$. We also take Boltzman's constant $k_B$ equal
to unity and $t = t_z = 1$ for all calculations and use  units where $\hbar=1,
e=1$. In addition, the lattice constant $a$ equals unity. In this regime, $t =
t_z$, there are four Weyl nodes in the first Brillouin zone of the
non-interacting Hamiltonian located at $(0, \pi, \pm \pi/2)$ and $(\pi, 0, \pm
\pi/2)$. Time-reversal symmetry is broken, but this model has other symmetries
detailed in Ref.~\onlinecite{Acheche_2019}. 

We introduce the orbital effects of a uniform magnetic field through the Peierls
substitution~\cite{Kohmoto1985}. The magnetic Hamiltonian (in the tight-binding form) then includes a
site-dependent Peierls phase $\phi_{nm}(B)$ that changes the symmetries of the
model. What about a Zeeman term? When the Pauli matrices act on spin, there is a Zeeman term whose effect on the Hamiltonian has been discussed in Ref.~\onlinecite{Acheche_2019}. In Appendix~\ref{AppendixZeeman} we show that in the non-interacting case the Zeeman term influences the magneto-optical spectrum only when lattice effects are important and we show that the effects are very small. The Landau wave functions are a combination of up and down spins and in the continuum limit, the Zeeman term simply shifts the position of the Weyl nodes and does not influence the spectrum at all. So from now on, we do not include a Zeeman term.   
We pick magnetic field values commensurate with the original lattice,
namely we take $eBa^2/h = p/q$ with rational values of
$p/q$~\cite{Hofstadter1976}. In the rest of this paper, and with no loss of
generality, we set $p=1$. The applied magnetic field is in the $z$-direction and
we use the Landau gauge (\ie a vector potential $\mathbf{A} = (0, Bx, 0$)) to
preserve translation symmetry along the $y$-direction. This leads to the
following Harper matrix:
\begin{equation} \label{Eq.HaperMatrix}
    \mathbf{H}_H = \begin{pmatrix}
    \mathbf{H}_{\uparrow\uparrow} & \mathbf{H}_{\uparrow\downarrow}\\
    \mathbf{H}_{\downarrow\uparrow} & \mathbf{H}_{\downarrow\downarrow}
    \end{pmatrix}.
\end{equation}
Each of the sub-matrices in the above equation is of dimension $q\times q$. They
are defined as follows,
\begin{subequations}
\begin{equation}
    \mathbf{H}_{\uparrow\uparrow} = 
    \renewcommand\arraystretch{1.8}
    \begin{pmatrix}
    M_0 & -t  & 0  & \dots & -t \\
    -t  & M_1 & -t & \dots & 0  \\
    0 & \ddots & \ddots & \ddots & \vdots\\
    -t & 0 & \dots & -t & M_{q-1}
    \end{pmatrix},\\
\end{equation}
\begin{equation}
    \mathbf{H}_{\uparrow\downarrow} = 
    \renewcommand\arraystretch{1.8}
    \begin{pmatrix}
    A_0 & -it  & 0  & \dots & it \\
    it  & A_1 & -it & \dots & 0  \\
    0 & \ddots & \ddots & \ddots & \vdots\\
    -it & 0 & \dots & it & A_{q-1}
    \end{pmatrix},
\end{equation}
\end{subequations}
with
\begin{equation}
\mathbf{H}_{\downarrow\uparrow} = \mathbf{H}_{\uparrow\downarrow}^{\dagger},\;\;\;\;
\mathbf{H}_{\downarrow\downarrow} = -\mathbf{H}_{\uparrow\uparrow},
\end{equation}
and with the definitions $M_n = -2t\cos{(k_y + 2\pi np/q)} - 2t_z\cos{k_z}$ and
$A_n  = 2it\sin{(k_y + 2\pi np/q)}$. 

Equation~\ref{Eq.HaperMatrix} describes a magnetic  unit cell with $q$ sites in the
$x$ direction. The full Hamiltonian includes a periodic extension of Harper
matrix along the $x$ axis, the chemical potential and the Hubbard interaction.
For the values of $q$ that we choose, the Harper matrix is sufficiently large
and the corresponding reduced Brillouin zone along the $k_x$ direction
sufficiently small that dependencies on $k_x$ can be neglected. These
dependencies are associated with the periodicity of the magnetic unit cell and
they become important in the large field limit where the magnetic unit cell has
only a few sites.  Consequently, the sites inside the magnetic unit cell have
identical local density of states (spin up + spin down) at half-filling.
Therefore, we are in the Landau regime defined in
Ref.~\onlinecite{Acheche_2017}.

In this approximation, the free Hamiltonian that takes maximum advantage of
translational invariance is
\begin{equation}
\hat{\mathbf{H}}_0 = \sum_{k_y, k_z} \hat{\mathbf{C}}^\dagger\mathbf{H}_H \hat{\mathbf{C}} - \mu \sum_i \hat{n}_i
\end{equation}
where the creation and destruction operators are defined in the basis:
$\hat{\mathbf{C}} = (\hat{c}_{1,k_y,k_z, \uparrow},\ldots \hat{c}_{q,k_y,k_z,
\downarrow})$.

We solve the interacting Hamiltonian non-perturbatively using the Dynamical Mean
Field Theory (DMFT) framework~\cite{Georges_1996}. The expected correction to
the DMFT self-energy~\cite{Schafer_2015} is an additive static momentum
dependent self-energy that would renormalize the energy dispersion and lead to
corresponding vertex corrections. These static corrections should not be
important when long-wavelength spin fluctuations are absent.
Ref.~\onlinecite{Acheche_2017} contains the derivation of the DMFT equations
that include the orbital effects of a uniform magnetic field. In summary, the
local self-energy depends on the magnetic field and the self-consistency
equation itself is unaltered. The effect of the magnetic field on the
self-energy comes from the non interacting density of states of the Landau levels
and the self-consistency equation.~\cite{Acheche_2017, Nourafkan_2014}

We use an exact diagonalisation (ED) impurity solver for the impurity problem
with a finite number of bath sites, $n_b$.~\cite{Caffarel_1994, Nourafkan_2011}
Though still of considerable size, the $n_b=5$ orbital Hamiltonian in this
scheme can be diagonalized exactly to compute the local Green’s function at
finite temperature.
 
\section{Magnetic field and electronic interaction effects on single-particle properties}\label{Sec:SinglePart}

We first consider the effects of magnetic field and interactions on the density
of states and then discuss the self-energy. This leads us to comment on the
resilience of quasiparticles in the presence of interactions in a Weyl
semimetal. 

\subsection{Density of states}
\begin{figure}
\includegraphics[scale=0.6]{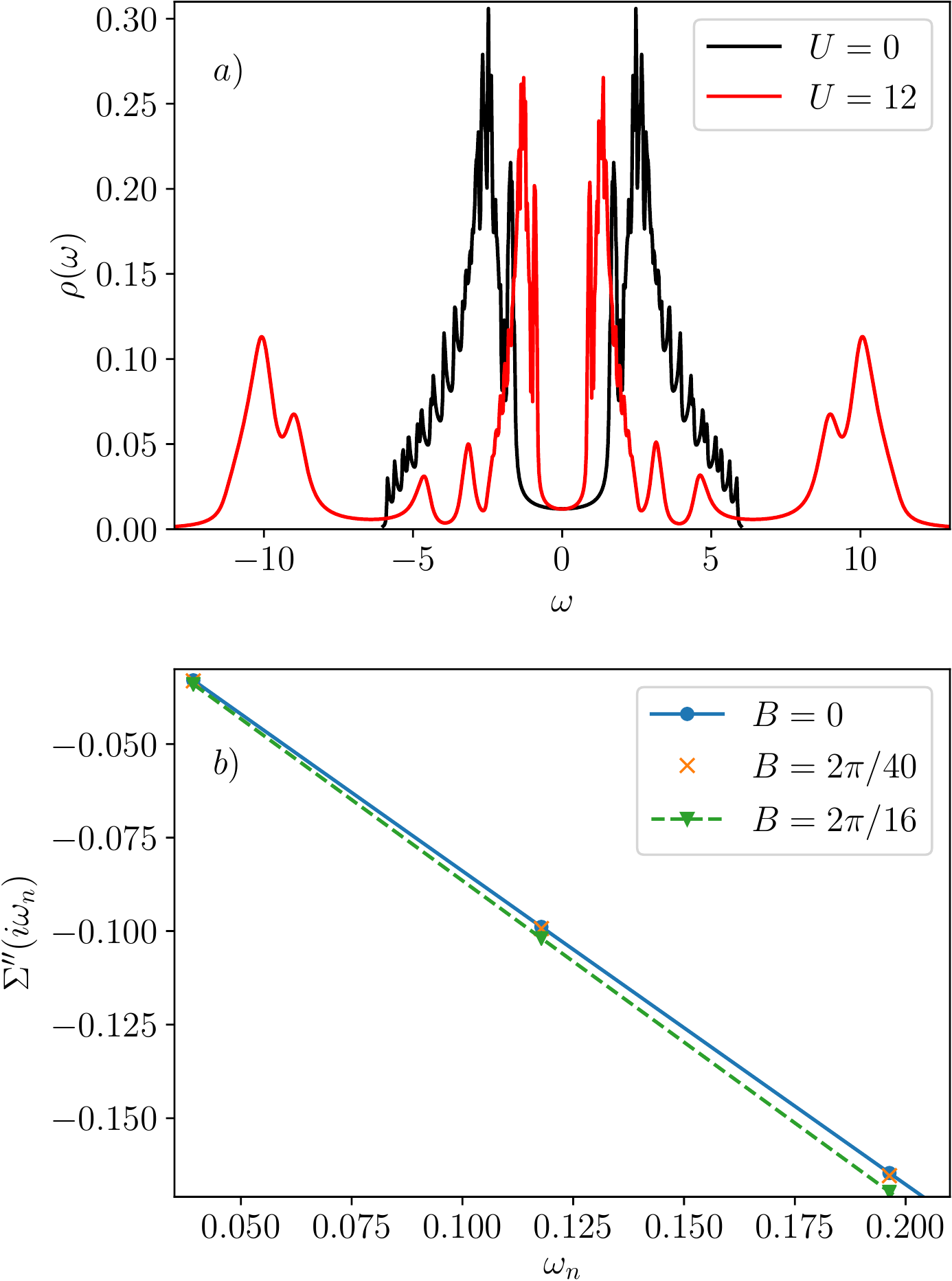}
\caption{(color online) a) Local density of states for noninteracting (black
line) and interacting (continuous red line) Weyl semimetals in the presence of an
external magnetic field ($B=2\pi/16$). A Lorentzian broadening $\eta=0.01$ has
been used for both plots. Features that are sharper than $0.01$ in frequency then, cannot be resolved. Note that the flat part of the spectrum near $\omega=0$ is insensitive to a wide range of values of $\eta$ since smoothing the density of states when it is already frequency independent does not change anything. 
 b) Imaginary part of the self-energy at
the three lowest Matsubara frequencies for the case without a magnetic field,
\ie $B=0$, the case where $B = 2\pi/40$, and finally in the quantum limit where
$B = 2\pi/16$. The inverse temperature is  $\beta = 80$.}\label{Fig.SelfDos}
\end{figure}

For a non-interacting Weyl semimetal without magnetic field, described by
Eq.~\ref{nonintHam}, the density of states  at low energy is characterized by
$\omega^2$ behavior, with a vanishing density of states  at $\omega=0$.  With a
magnetic field, a finite field-dependent density of states  appears at low
energies as shown in Fig.~\ref{Fig.SelfDos}a. The non-interacting density of
states  in this figure is obtained from 
\begin{equation}
    \rho(\omega) = \frac{1}{N}  \sum_n \int \frac{\text{d}k_y \text{d}k_z}{(2\pi)^2} \delta (\epsilon_{n, k_y, k_z} - \omega)
\end{equation}
where $N$ is the total number of Landau levels,  $n$ is the Landau level index
and $\epsilon_{n, k_y, k_z}$ the corresponding dispersion energy obtained from
the Harper Matrix. The finite field-dependent density of states at low energy
increases with increasing field.  Apart from this, a magnetic field does not
influence the main characteristics and the bandwidth of the density of states of
a non-interacting Weyl semimetal.  

Fig.~\ref{Fig.SelfDos}a also shows the density of states of the interacting
system. Apart from the lower and upper Hubbard bands (incoherent satellites
around $\omega=\pm10$ on Fig.~\ref{Fig.SelfDos}a), the interaction tends to
shrink the coherent quasiparticle bandwidth by the quasiparticle weight $Z$.
However, the density of states at the Fermi level is not affected by electronic
correlations. This can be explained as follows. 
Physically, the spectral function is proportional to the quasiparticle weight
$Z$, but the one-dimensional density of states of the chiral level is
proportional to one over the renormalized Fermi velocity $Zv_F$ so that the two
factors of $Z$ cancel each other. 
The density of states for chiral Landau levels in the quasipaticle approximation
is then given by
\begin{equation} \label{Eq.DOSCLL}
\rho(\omega) \propto \frac{\Theta \left[Z^{-1}(\omega/ v_F )+ k_c \right] - \Theta \left[Z^{-1}(\omega/v_F) - k_c \right]}{|v_F|},
\end{equation}
where $v_F$ stands for the Fermi velocity, $k_c$ is an energy cut-off and
$\Theta$ is the step function. Equation~\ref{Eq.DOSCLL} shows that the density of
states is insensitive to the quasiparticle spectral weight $Z$, but the range of
frequencies where it applies is narrowed down. 
There is another way to explain that the density of states at the Fermi level is
independent of interactions: It is a general property of single-site DMFT at low
enough temperature. Indeed, one can show, using Luttinger's theorem for a
momentum-independent self-energy~\cite{Muller_1989}, that the density of states
at the Fermi level ($\omega=0$) is independent of interactions. One has to
assume that this remains valid for a range of energies close to the Fermi level.

\subsection{Self-energy}
The scattering amplitude is related to the imaginary part of the self-energy,
which depends on $U$. At fixed $U$, we investigate the effect of the magnetic
field on the imaginary part of the self-energy at low temperature (corresponding
to an inverse temperature $\beta = 80$) by comparing the case without a magnetic
field, \ie $B=0$, with the case where $B=2\pi/40$, and with the case where
$B=2\pi/16$, \ie in the quantum limit. The latter is characterized by a clear
separation between the zeroth and the first non-zero Landau levels.

Figure~\ref{Fig.SelfDos}b illustrates the effect of magnetic field on the
imaginary part of the self-energy for the first few Matsubara frequencies at
$\beta = 80$, $U = 12$ and half-filling. As in the case of the Hubbard model on
the square lattice~\cite{Acheche_2019}, the magnetic field does not have a large
effect on the self-energy. For $B=2\pi/ 40$, there are few differences with the
self-energy without a magnetic field. However, for $B = 2\pi/16$ we can clearly
see that the self-energies depart from each other at larger Matsubara
frequencies.
    
Those differences at intermediate frequencies come from the significant
modification of the local density of states due to the orbital effect of the
magnetic field. It is also interesting to note that despite these differences,
the three self-energies have the same value at the first Matsubara frequency.
This indicates that the scattering times at the Fermi level for half-filling are
essentially the same with or without magnetic field.
    
\subsection{Resilience of quaiparticles}
 
Although it is not completely apparent from the above results, quasiparticles
are remarkably resilient in a Weyl semimetal. To show this, let us momentarily
remove the magnetic field. 
%
At small interaction strength, the self-energy can be calculated with the iterated perturbation theory (IPT)
solver \cite{kajueter_1996_IPT} but without the DMFT self-consistency. Then one
can use the quadratic effective density of state of Weyl semimetals near the
Fermi energy and obtain analytically the imaginary part of the self-energy at
low frequency for a single Weyl node (see the derivation in
Appendix~\ref{Appendix1}): 
\begin{equation}\label{Eq.selfpert}
\Sigma''(\omega) \propto -\frac{U^2 \omega^8}{\pi^5 v_F^9},
\end{equation}
where $v_F$ is the Fermi velocity. Nevertheless, one expects that since this
suggests the existence of quasiparticles, this result should be valid for larger
values of $U$. Note that this unusual behavior is very different from Fermi
liquid theory where $\Sigma''(\omega)$ is quadratic in frequency. From the point
of view of lifetime, the ``Weyl liquid'' seems to lead to even more stable
quasiparticles than Fermi liquids. 

But what about the quasiparticle spectral weight $Z$? Appendix~\ref{Appendix1}
shows that there is a finite value of $Z$ and calculations show that, as
expected, it decreases (roughly as $-U^2$) as $U$ increases. The smallness of
the imaginary part of the self-energy at low frequency offers a clue for the
robustness of quasiparticle physics. This justifies a quasiparticle approach
where the main effect of the interactions is encoded in the quasiparticle weight
with an (extermely) small lifetime for the quasiparticles. However, the above
derivation is only valid in the absence of magnetic field since it relies on a
vanishing density of states at the Fermi level. When an external magnetic field
is applied, a finite density of states appears and that could change the
physics.

\section{Conductivity of interacting Weyl Semimetals}\label{Sec:Cond}

In this section, we use our knowledge of single-particle properties to compute
the optical conductivity and find out how it is affected by magnetic field,
temperature and finite bandwidth. 

Let us first consider the current-current correlation function in Matsubara
frequency. In linear response theory at zero momentum and neglecting vertex
corrections, this correlation function $\Pi_{zz}(\mathbf{q}\rightarrow 0,
i\nu_n)$ along the $z$ direction is~\cite{Mahan, Nourafkan_2018}
\begin{align} 
\Pi_{zz}(i\nu_n)& =- \frac{\pi }{N \beta} \sum_{\mathbf{k}\omega_m} \Tr \bigg[\mathbf{a}_{zz}(\mathbf{k}) \mathbf{G}(\mathbf{k}, i\omega_m)  \nonumber\\
 &-\mathbf{v}_z(\mathbf{k}) \mathbf{G}(\mathbf{k}, i\omega_m+i\nu_n) \mathbf{v}_{z}(\mathbf{k})\mathbf{G}(\mathbf{k}, i\omega_m) \bigg]\label{Eq.Pola},
\end{align} 
where $\beta = 1/(k_BT)$ is the inverse of temperature, $i\omega_n$ is the
fermionic Matsubara frequency, $\Tr$ is the trace over $2q\times2q$ matrices,
$\mathbf{a}_{zz}(\mathbf{k})= \partial^2_{k_z} \mathbf{H}_H$ is the inverse
effective mass tensor, $\mathbf{v}_z(\mathbf{k})=\partial_{{k_z}}\mathbf{H}_H$
the velocity matrix along the $z$ direction, and $\mathbf{G}(\mathbf{k},
i\omega_m)$ is the interacting Matsubara Green function. $\Pi_{zz}$ is composed
of two parts: $\Pi^{Dia}$, the diamagnetic part and $\Pi^{Para}$, the
paramagnetic part (respectively, first and second term in Eq.~\ref{Eq.Pola}).
Gauge invariance imposes that these two parts cancel each other perfectly at
$\nu_n = 0$. After analytic continuation, the real part of the retarded
conductivity $\sigma'_{zz}$ is
\begin{equation}\label{Eq.Firstexpressionsigma}
\sigma'_{zz}(\omega) = \frac{\Pi''_{zz}(\omega)}{\omega},
\end{equation}
where $\Pi''_{zz}(\omega)$ is the imaginary part of $\Pi_{zz}(\omega)$. One can
also compute the real part of conductivity directly in real frequency
from~\cite{Nourafkan_2012}
\begin{align}\label{Eq.conductivity}
\sigma'_{zz}(\omega) = &\frac{\pi }{N} \sum_{\mathbf{k}} \int \text{d}\epsilon \Tr\bigg[ \mathbf{v}_z(\mathbf{k})  \mathbf{A}(\mathbf{k},\epsilon) \mathbf{v}_z(\mathbf{k})  \mathbf{A}(\mathbf{k},\omega +\epsilon)\bigg] \nonumber \\&\times \frac{f(\epsilon) - f(\omega+\epsilon)}{\omega},
\end{align}
where $\mathbf{A}$ is the spectral function matrix and $f$ is the Fermi-Dirac
distribution function. In the DC limit, the difference of the Fermi-Dirac
distribution functions in Eq.~\ref{Eq.conductivity} can be replaced by the
derivative with respect to frequency at $\omega=0$. Equation~\ref{Eq.conductivity} is
valid only when the trace is real, otherwise, additional terms coming from the
paramagnetic term of the current-current correlation function have to be taken
into account to obtain the real part of the conductivity. Moreover, since the
system breaks time-reversal symmetry, the spectral weight must be computed from
the anti-hermitian part of the Green's function matrix
\begin{equation}
\mathbf{A}(\mathbf{k},\omega) = \frac{1}{2i\pi}\left[\mathbf{G}^\dagger(\mathbf{k}, \omega) - \mathbf{G}(\mathbf{k}, \omega) \right],
\end{equation}
where the spectral weight is normalized as follows, $\int d\omega
\mathbf{A}(\mathbf{k},\omega)=\mathbf{1}$. The real-frequency dependent Green's
functions are obtained by analytic continuation using the Padé approximant
method~\cite{Vidberg1977} with a Lorentzian broadening $eta=0.01$.

\subsection{Interaction effects in the low-temperature limit}
\label{subsec:opticalcond}

\begin{figure}
\includegraphics[scale=0.55]{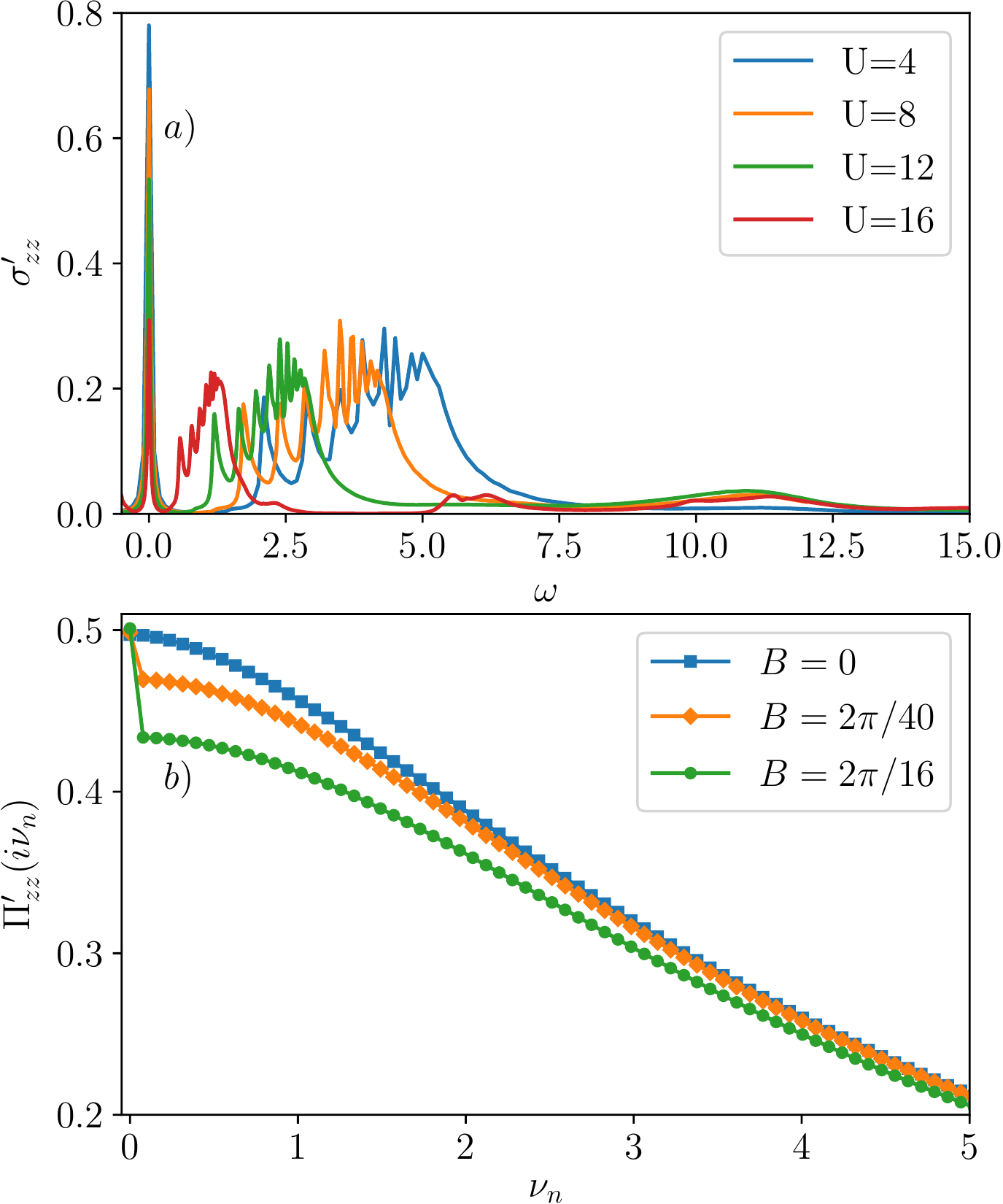}
\caption{(color online) a) Magneto-optical conductivity of the interacting Weyl
semimetal in an external magnetic field for several interaction strengths and
$\beta=80$.  b) Magneto-optical polarization as a function of bosonic Matsubara
frequency for $B=2\pi/16$, $B=2\pi/40$ and $B=0$, (\ie without external magnetic
field in the latter case).  DC conductivity scales with the jump at the lowest
frequency. The figure shows that DC conductivity (resistivity) is increasing
(decreasing) upon increasing field, \ie, reducing $q$.}\label{Fig.opticalcond}
\end{figure}

Figure~\ref{Fig.opticalcond} shows the magneto-optical conductivity of the
interacting Weyl semimetal for several interaction strengths and $B=2\pi/16$. It
is an even function of frequency. In the presence of Hubbard interactions, the
magneto-optical conductivity has three well defined features: The Drude Peak
near zero frequency, the interband transitions between Landau
levels~\cite{Ashby2013a} and incoherents peaks that are a consequence of Hubbard
bands in the density of states. The interband transitions at lower frequency are
quite similar to the results obtained from a non-interacting continuum model of
Weyl nodes~\cite{Ashby2013a}, \ie a series of asymmetric peaks superimposed on
the linear background from the no-field case. In our case, the lattice
introduces a natural cutoff and differences with the continuum model at higher
energy.   

The three parts of the optical conductivity are affected differently by the
Hubbard interaction. It is the chiral zero'th Landau level that leads to a
finite density of states at the Fermi level and to a Drude peak in the optical
conductivity.~\cite{Ashcroft, Son_2013} Upon increasing $U$, the Drude peak
decreases in intensity and in weight because of the quasiparticle weight $Z$,
even if the density of state at $\omega=0$ is not affected by electron-electron
interactions. The optical spectral weight of the interband contribution is also
reduced by interaction and the optical weight is transferred to the incoherent
satellite at high energy. Furthermore, the optical gap between low-frequency
peak and the interband contribution is decreased by the interaction, as can be
seen from Fig.~\ref{Fig.opticalcond}a.

A quantitative study of the Drude peak in real frequency is cumbersome since it
requires analytic continuation of numerical data. This could introduce
artifacts, especially in presence of interactions. Fortunately, at low
temperatures the weight of the Drude peak can be extracted from the
Matsubara-frequency current-current correlation function
$\Pi^{Para}_{zz}(i\nu_n)$. As shown in Fig.~\ref{Fig.opticalcond}b,
$\Pi^{Para}_{zz}(i\nu_n)$ is a smooth function of the Matsubara frequencies
except at the lowest frequency, \ie, $\nu_0=0$. The sudden increase at this
frequency indicates a non-zero DC conductivity. Indeed, in an insulator,
$\Pi^{Para}_{zz}(i\nu_n)$ smoothly reaches its zero frequency value as one can
see from $B=0$ result. Furthermore, in the interacting Weyl semimetals at low
temperatures, the Drude peak is separated from the rest of the optical spectrum
by a clear gap (see Fig.~\ref{Fig.opticalcond}a). This  allows us to show that
the jump in $\Pi^{Para}_{zz}(i\nu_n)$ at the lowest frequency scales with the
Drude peak weight. The details of the derivation are presented in
Appendix~\ref{Appendix}.

\begin{figure}
\includegraphics[scale=0.55]{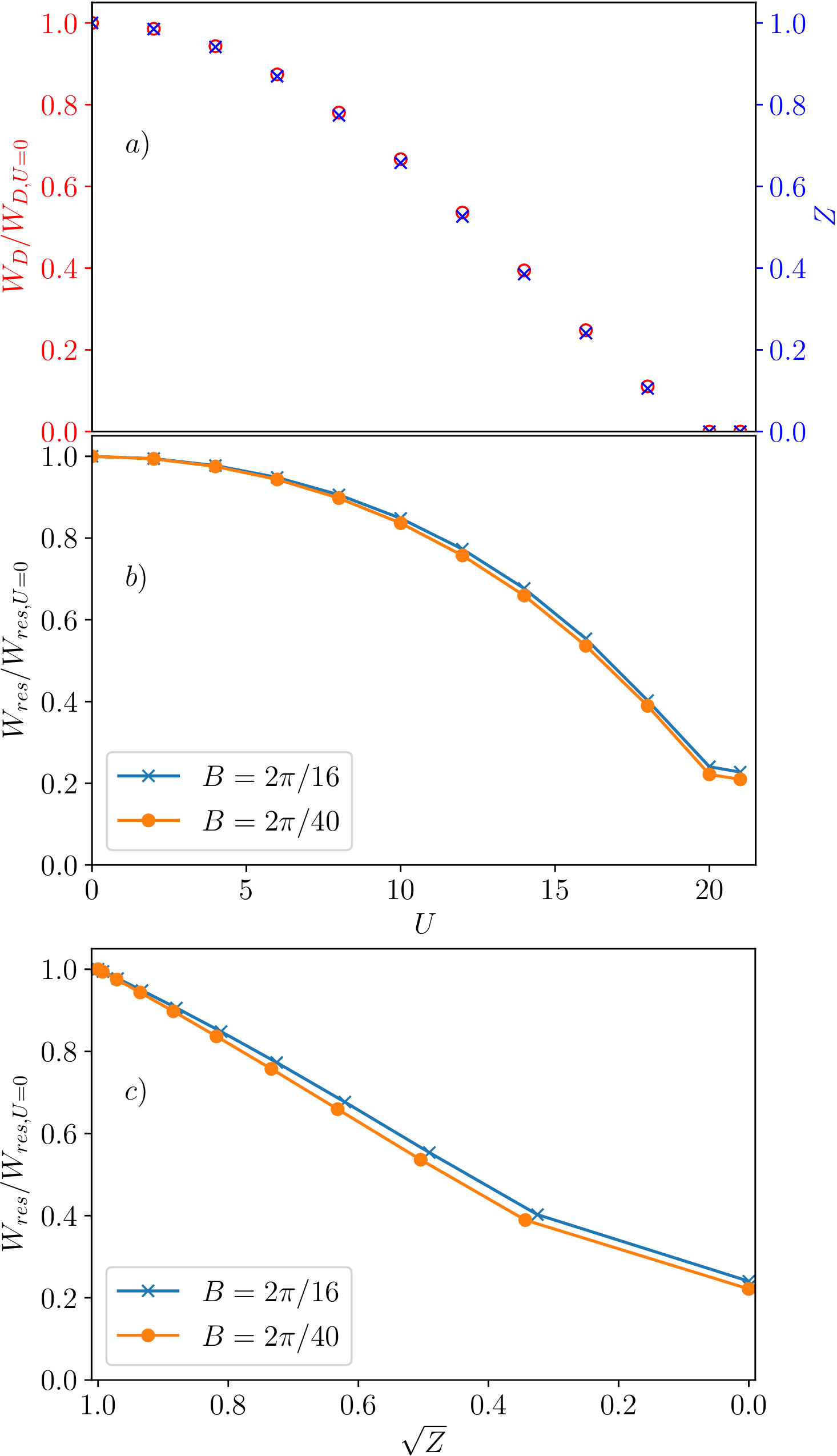}
\caption{(color online) a) Interacting Drude-peak weight normalized by the
non-interacting one (left vertical axis) and quasi-particle weight for
$B=2\pi/16$ (right vertical axis) both as a function of interaction strength.
Both quantities show identical interaction dependence.  b) Spectral weight of
the Drude peak normalized by the interband spectral weight as a function of $U$,
calculated using Eq.~\ref{Eq.ratio}. c) Normalized $W_{res}$  as a function of
square root of the quasiparticle weight $\sqrt{Z}$ that suggests $Z^{1/2}$
dependence. For all panels, $\beta=80$.}\label{Fig.DrudeWeight}
\end{figure}

With the help of Eq.~\ref{Eq.DrudeWeightEq}, we are able to follow the fate of
the Drude weight, $W_D=\int \text{d}\omega \sigma^D(\omega)$, with and without
interaction.  Here, we define
$\sigma^{\prime}_{zz}(\omega)=\sigma^D_{zz}(\omega)+\sigma^{res}(\omega)$
because  the Drude peak is separated with an energy gap from the rest of the
optical spectrum. As one can see from Fig.~\ref{Fig.DrudeWeight}a, at low
temperatures the interaction dependence of the  low-frequency peak weight
normalized by the non-interacting one follows exactly the quasiparticle weight
for all values of $U$ tested. The normalized Drude weight scales like the
quasiparticle weight $Z$, even near the Mott transition~\cite{morimoto2016weyl} (around $U=20$), a clear
sign of the robustness of quasiparticle physics. 

The shrinking of interband contributions with increasing $U$ is also a sign of
quasiparticle physics. Indeed, the quasiparticle weight $Z$ renormalizes the
whole band, so the Landau levels become closer to each other. This leads to
excitations at lower frequency than in the absence of interactions. Contrary to
interband transitions, transitions between incoherent Hubbard bands and
quasiparticle bands and between the lower and the upper Hubbard bands is due to
the frequency dependence of the self-energy and cannot be explained by the
quasiparticle spectral weight alone. However, $Z$ still governs a large part of
the optical conductivity. 

In Fig.~\ref{Fig.DrudeWeight}b, we present  $W_{res}=\int \text{d}\omega
\sigma^{res}(\omega)$ normalized by its value at $U=0$.  This quantity is easily
computed with the help of Eq.~\ref{Eq.DrudeWeightEq} and of
\begin{equation} \label{Eq.ratio}
W_{res}=\int^{+\infty}_{-\infty} \text{d}\omega \sigma^{res}(\omega) = {\tilde{\Pi}^{Para}_{zz}(i\nu_n=0)},
\end{equation}
where $\tilde{\Pi}^{Para}_{zz}$ denotes the paramagnetic part of the
current-current correlation function without the jump at the lowest frequency.
It can be obtained from an extrapolation to $\nu_n=0$ of a polynomial fit of
$\Pi^{Para}_{zz}(i\nu_n\neq 0)$. 

In the weak to intermediate range of interaction, $W_{res}$ depends only weakly
on the interaction strength.  For interaction strengths larger than the bare
band-width,  $W_{res}$ decreases at a faster rate and eventually saturates to a
finite value when the system undergoes a phase transition to a Mott insulator.
Hence the variation of $W_D$ or of the effective mass with interaction is more
pronounced than the variation of $W_{res}$.  

Figure~\ref{Fig.DrudeWeight}c shows the normalized  $W_{res}$  as a function of
square root of the quasiparticle weight $\sqrt{Z}$. The point at $Z=0$ is in the
insulator. The linearity of the curve and the value of the slope, equal to $1/2$
over the whole range, except for the transition from metal to insulator,
indicates that the ratio is directly proportional to the square root of $Z$. We
have not found a simple argument for this result.  

\begin{figure}[h!]
\includegraphics[scale=0.6]{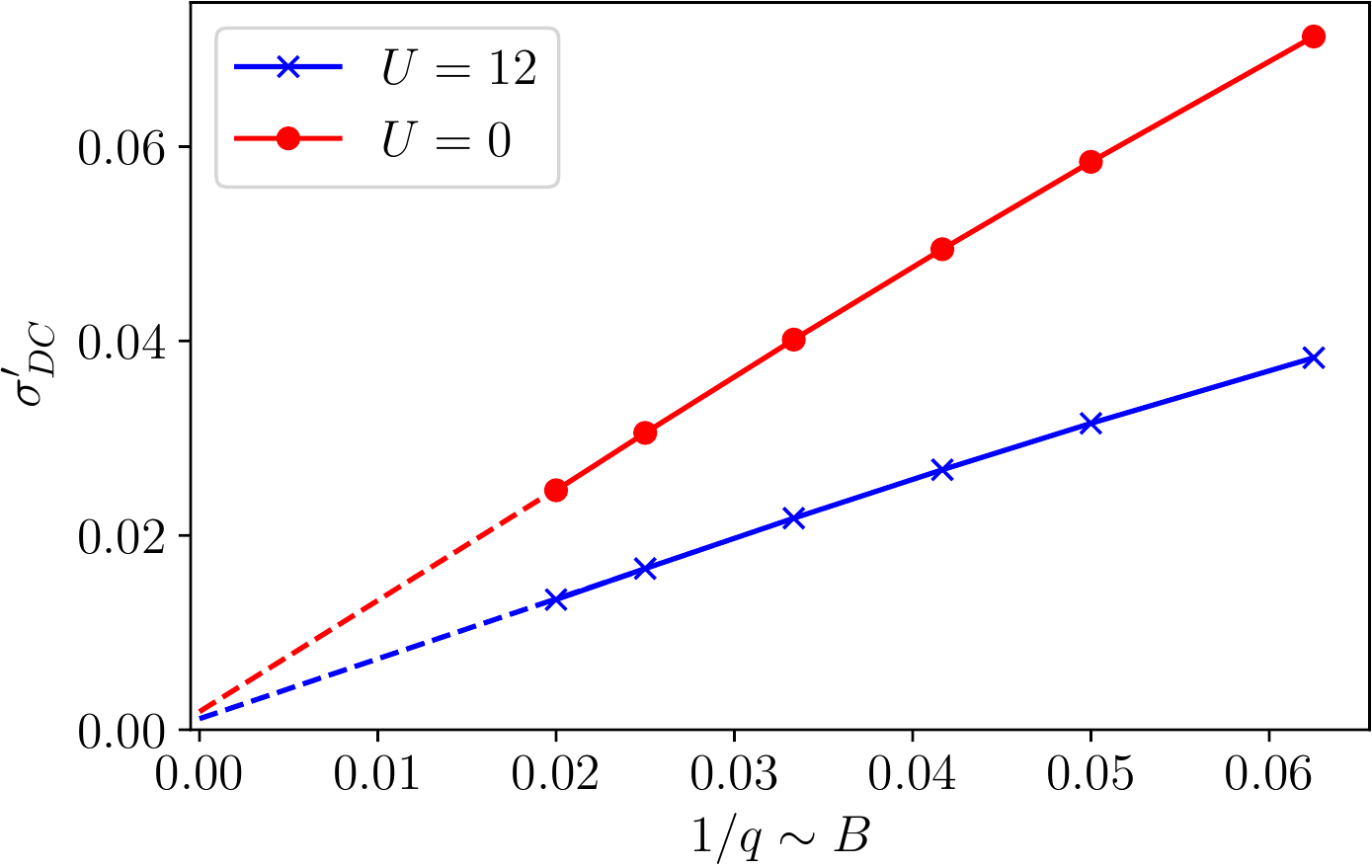}
\caption{DC conductivity of an interacting Weyl semimetal at $\beta=80$ as a
function of the external magnetic field.  Conductivity is linearly increasing
upon increasing the field strength. The slope depends on the interaction strengh
and decreases when $U$ increases. The dashed parts are linear extrapolation
based on the first three values of the DC conductivity.}\label{Fig.DCcond}
\end{figure}

Finally, consider the magnetoresistance of a Weyl liquid. For the values of $q$
tested in this paper ($ q \in [16, 100 ]$), the Drude weight increases linearly
with the magnetic field (see figure~\ref{Fig.DCcond}). This is the famous
negative magnetoresistance phenomenon, which is a consequence of the quantum
limit where only the chiral Landau levels contribute to the Drude
peak~\cite{Son_2013}. As one can see from Fig.~\ref{Fig.DCcond}, the linear
dependence of the conductivity is not impacted by the interaction but the slope
decreases upon increasing $U$.  At higher $U$, the system undergoes a phase
transition to a Mott phase with zero DC conductivity. Electron-electron
interactions do not destroy the negative magnetoresistance, they only
renormalize it, as expected from the quasiparticle picture. This statement finds
its experimental proof since Weyl physics has been observed in correlated
materials.~\cite{Kuroda2017}

\subsection{Temperature effects}
\begin{figure}[h!]
    \includegraphics[scale=0.35]{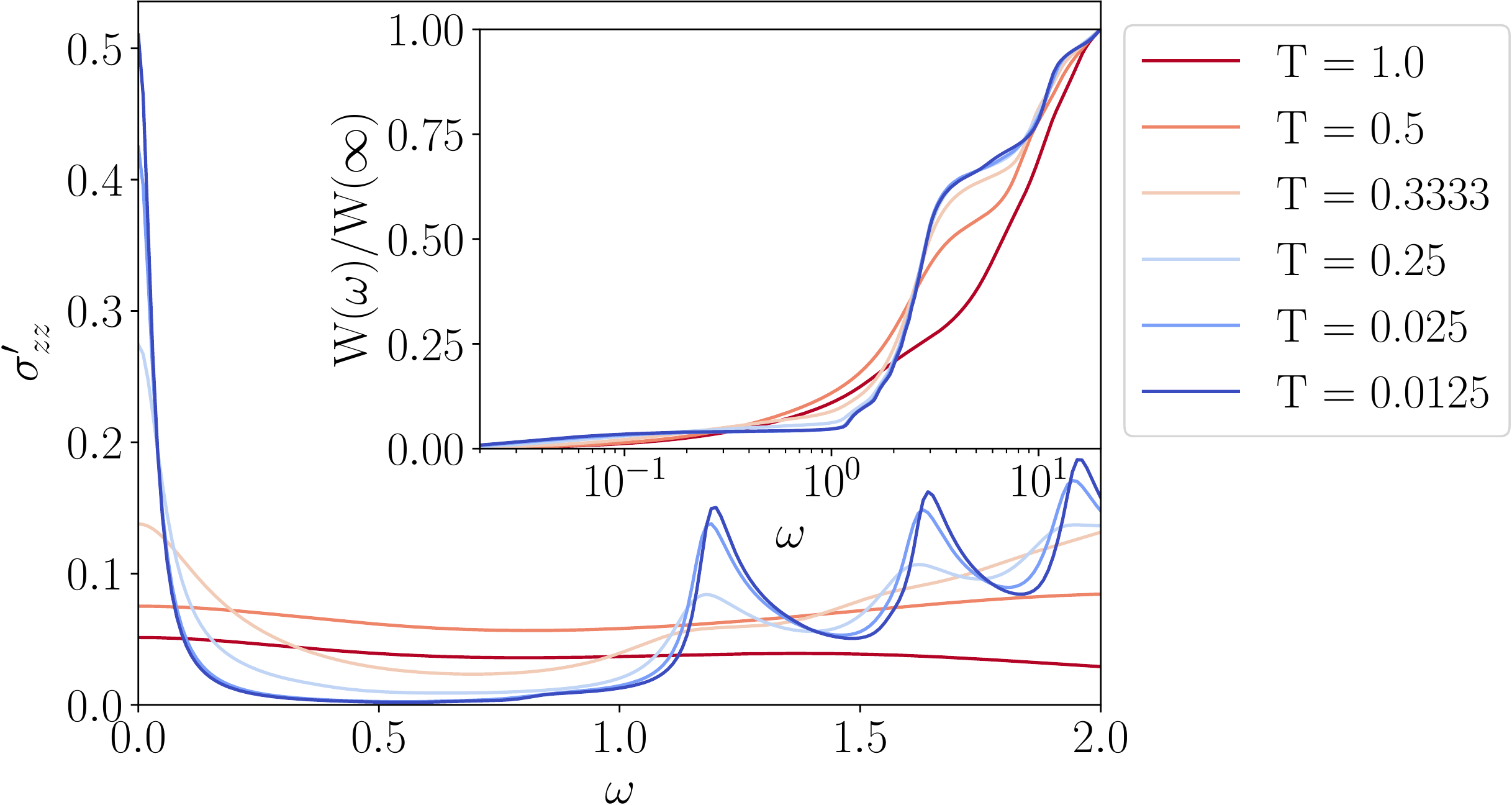}
    \caption{(color online) Magneto-optical conductivity of an interacting Weyl
    semimetal for several temperatures \shaheen{at $B= 2\pi/40$ and $U=12$.} The
    inset shows the magneto-optical spectral weight $W(\omega)$ as a function of
    cutoff frequency (cf. Eq.~\ref{Eq.spectralweight}). Three distinct plateaux
    corresponding to the three features discussed in~\ref{subsec:opticalcond}
    can be seen in the inset}\label{Fig.OpticalTemperature}
    \end{figure}

    \begin{figure}[h!]
        \includegraphics[scale=0.40]{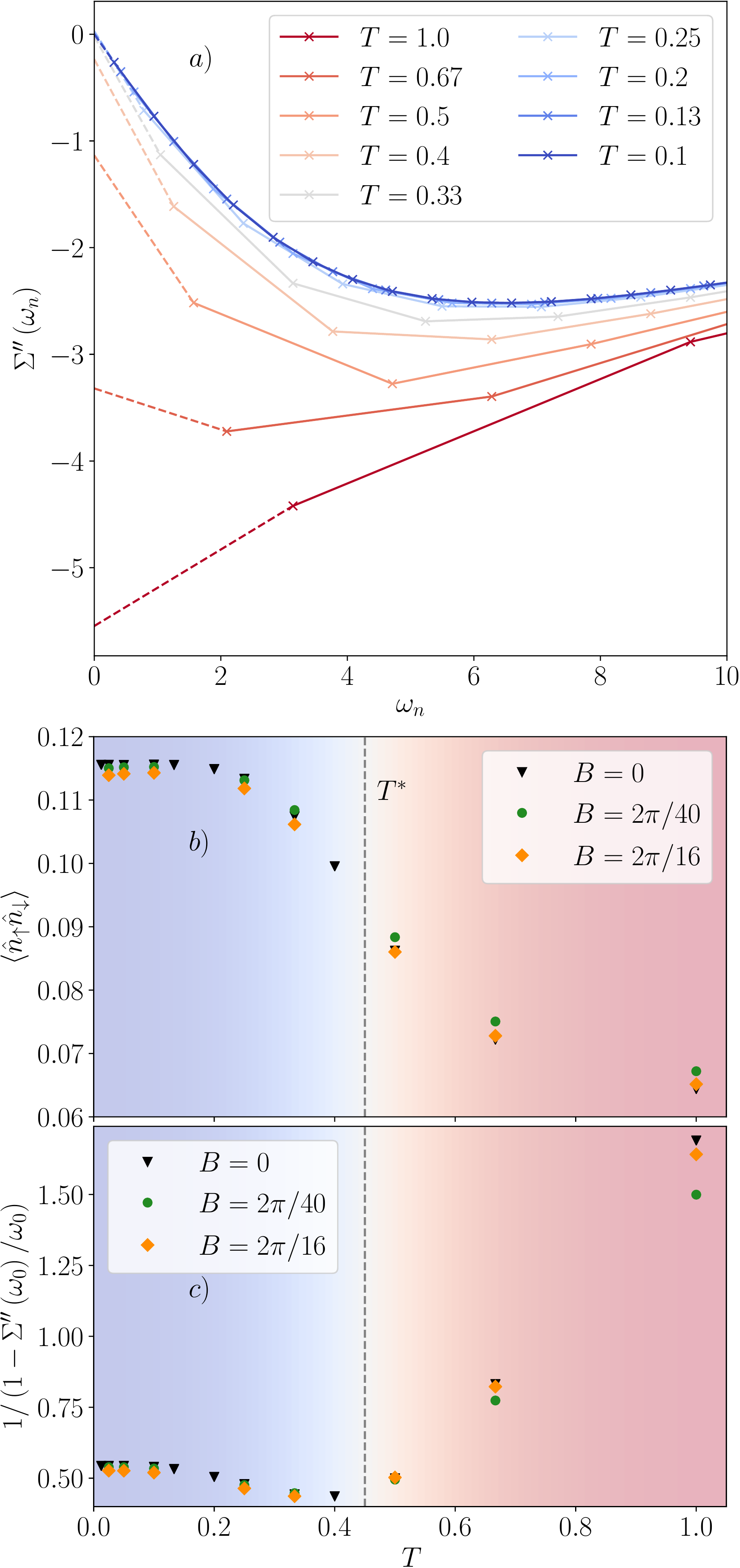}
        \caption{(color online) a) Imaginary part of the self-energy as a
        function of Matsubara frequencies for different temperatures. $U$ is
        fixed to $12$ and the system is at half-filling. The continuous lines
        are computed using DMFT. The dashed lines are a zero frequency
        extrapolation of the self-energies based on a fourth-order polynomial
        fit of the first five values of the self-energy. At low temperature, the
        self-energy extrapolates to zero like in ordinary metals. However, at
        higher temperature, it extrapolates to a value far from being zero at
        zero frequency. b) Double occupancy as a function of temperature for
        $U=12$ and for several magnetic fields. $T^*$ can be defined by the
        inflection point around $0.45$, marked by a vertical dashed line. c)
        $T^*$ coincides with the minimum in the approximation $(1-{\rm
        Im}\Sigma(\omega_0)/\omega_0)^{-1}$ for the single-particle spectral
        weight $Z$. Values of $(1-{\rm Im}\Sigma(\omega_0)/\omega_0)^{-1}$
        larger than unity are not physical and in fact quasiparticles disappear
        in the regime where there is an increase after reaching a minimum.}
        \label{Fig.DoubleOcc}
    \end{figure}
Bad metal behavior is a consequence of interactions.~\cite{Brown_2019,
Deng_2013, Hussey_2004} In a Fermi liquid (FL) the crossover between a coherent
metal and a bad metal can be identified from the frequency dependence of the
optical conductivity at low frequencies~\cite{Deng_2013}. At low-$T$, the Drude
peak of a FL decays as $1/\omega^2$. The crossover out of the FL regime leads to
a broader low-frequency peak whose frequency dependence is no longer
$1/\omega^2$. At the transition to the bad metal regime the Drude and interband
features merge.~\cite{Deng_2013} Similar behavior can be seen in the optical
conductivity of a Weyl semimetal. At low-$T$, the Drude peak is separated from
the interband contribution and it decays very quickly with frequency. At higher
temperatures though, both Drude peak and interband contributions broaden and
merge together as one can see from Fig.~\ref{Fig.OpticalTemperature} for large
interaction, here $U=12$ equal to the non-interating bandwidth. 

In fact, the merging of the Drude peak with interband transitions is just one of
the manifestations of what happens on much larger energy scales. Indeed, the
crossover affects the optical conductivity on all frequency scales. To
illustrate this, we calculated the optical spectral weight integrated up to a
cutoff $\omega$
\begin{equation}\label{Eq.spectralweight}
W(\omega) = \int^\omega_0 \text{d}\nu\; \sigma'_{zz}(\nu),
\end{equation}
and we plotted it as an inset in Fig.~\ref{Fig.OpticalTemperature}. At low
temperatures, three frequency ranges can be identified: the Drude weight for
$\omega \in [0, 1]$ followed by the intraband contributions and finally the
Hubbard band contributions at higher energies. This three-part structure is
unchanged until the crossover between metal and bad metal occurs at $T^*$. Above
$T^*$, the Drude and finite frequency features merge. That temperature affects
dynamical properties over scales much larger than thermal energies is
characteristic of strongly correlated systems.

In bad metals, the quasi-particles become ill-defined. This can be seen from the
quasi-particles scattering rate, given by the imaginary part of the self-energy
on the real axis. At low enough temperature, it can be approximated by the
imaginary part of the Matsubara self-energy using
\begin{equation}
    Z\approx \left(1 - \frac{\Sigma''(\omega_0)}{\omega_0} \right)^{-1}.
\end{equation}
In Fig.~\ref{Fig.DoubleOcc}a, the continuous lines represent the value of the
self-energy obtained from DMFT for $U=12$ and at different temperatures. The
dashed lines represent a zero frequency extrapolation based on a fourth order
polynomial fit of the first five values of the self-energy. It extrapolates to
zero frequency at low temperature, consistent with existence of the low-energy
well-defined excitations, but at higher temperature this behavior is no longer
observed. 
    
The crossover temperature can be identified clearly from static quantities as
well. Consider double occupancy, shown in Fig.~\ref{Fig.DoubleOcc}b. The
crossover temperature is around the inflection point of these curves at
$T^*\approx 0.45$, consistent with the optical conductivity. Moreover, we also
find that the crossover temperature in the presence of a magnetic field is very
close to the crossover temperature obtained without magnetic field. This is
because, as shown for example in Ref.~\onlinecite{Deng_2013}, the bandwidth and
$U$ are  clearly the two energies that control $T^*$. The energies associated
with the magnetic fields that we consider are small compared with the bandwidth
and with $U$, and are thus not very relevant. 

The crossover can also be seen from the temperature dependence of $(1-{\rm
Im}\Sigma(\omega_0)/\omega_0)^{-1}$, plotted at the lower panel of the
Fig.~\ref{Fig.DoubleOcc}c. At low-$T$, this quantity gives the quasi-particle
$Z$. As one can see, it reaches a minimum around $T^*\approx 0.45$ and becomes
even larger than unity at high enough temperature. This behavior is not physical
and coincides with the disappearance of quasi-particles.

\section{Conclusion}\label{Sec:Conclusion}
In summary, our study reveals that an interacting Weyl semimetal is extremely
robust to short-range interactions. The density of states at the Fermi level
coming from the magnetic-field-induced chiral level is not modified by
interactions. Interactions are manifest mostly through a quasiparticle
renormalization $Z$, as in a Fermi liquid, but with a frequency-dependent
self-energy that vanishes much faster with frequency as one approaches the Fermi
level. The slope of the negative magnetoresistance dependence on the field is
reduced by $Z$. At elevated temperatures, Weyl liquids exhibit a crossover to a
bad metal phase at a crossover temperature that is essentially magnetic-field
independent.  

\acknowledgments

This work has been supported by the Natural Sciences and Engineering Research
Council of Canada (NSERC) under Grant No. RGPIN-2014-04584, by the Research Chair in
the Theory of Quantum Materials, by the Canada First Research Excellence Fund
and by the Canadian Institute for Advanced Research. Simulations were performed
on computers provided by the Canadian Foundation for Innovation, the Minist\`ere
de l'\'Education des Loisirs et du Sport (Qu\'ebec), Calcul Qu\'ebec, and
Compute Canada.

\appendix

\section{Effect of the Zeeman term on the magneto-optical conductivity}\label{AppendixZeeman}

In this Appendix, we show that even unrealistically large values of the Zeeman term $h$ do not appreciably change the magneto-optical spectrum, even though the Landau levels can be strongly modified. Fundamentally, this comes from selection rules and correlated changes of the eigenenergies in the valence and conduction bands. 

We start with the weak magnetic-field case $B=2\pi/40$ and then consider in more detail a stronger field, $B=2\pi/16$ where lattice effects are more apparent and modifications of the magneto-optical conductivity more noticeable. We close by studying the case where the Zeeman term is so strong that there is a topological transition from four to two Weyl nodes. Even though we restrict ourselves to the non-interacting case, it suffices to keep in mind the quasiparticle picture, Hubbard bands and Mott transition to guess the qualitative effects of interactions.  

\begin{figure}[h!]
\centering
      \centering\includegraphics[width=\linewidth]{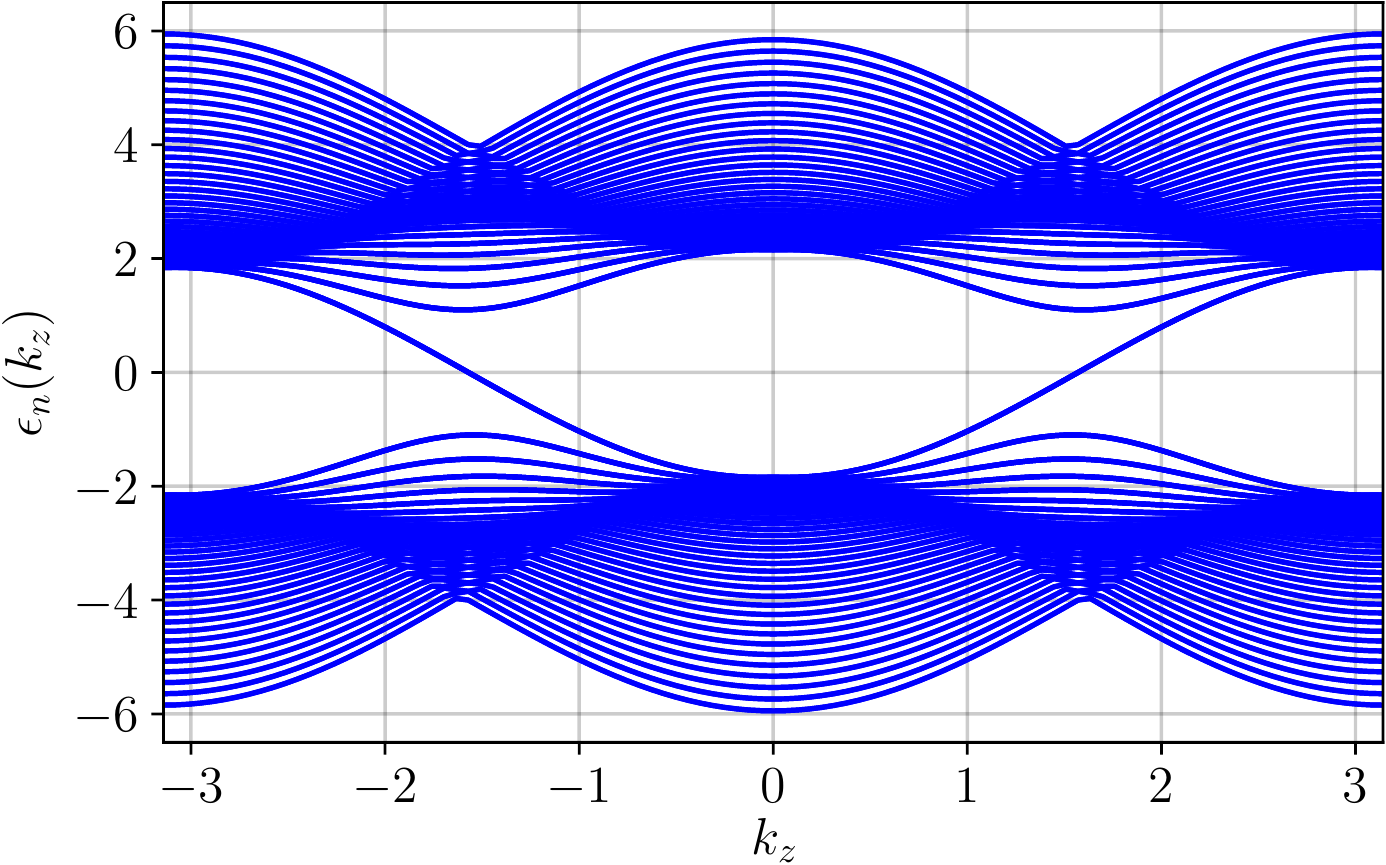}
      \caption{Landau levels for $B=2\pi/40$, $h=0$}
      \label{Fig.Levels40}
\end{figure}

\begin{figure}[h!]
      \centering\includegraphics[width=\linewidth]{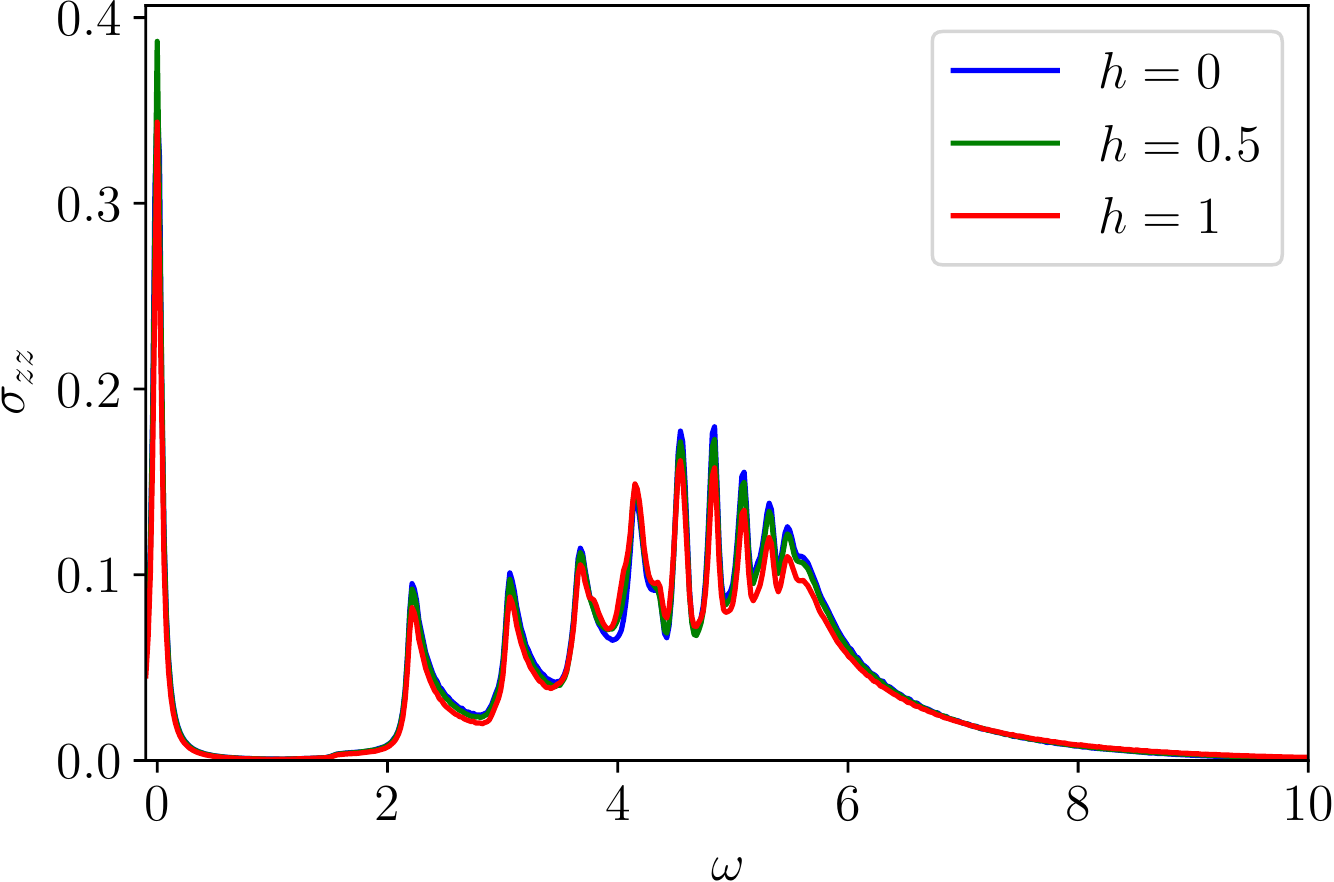}
      \caption{Magneto-optical spectrum for $B=2\pi/40$, $h=0, 0.5, 1$. It is only at the highest energies that small changes can be detected. In the low-energy limit, up to $\omega\sim 4$, one can understand the lack to $h$ dependence from the continuum model.} 
      \label{Fig.Conductivity40}     
\end{figure}

It has been shown in Ref.~\onlinecite{Acheche_2019} that, for our model, the Weyl nodes remain at $\mu=0$ even when $h$ differs from zero. The Landau levels for $B=2\pi/40$ and $h=0$ appear in Fig.~\ref{Fig.Levels40}. The corresponding magneto-optical conductivity for three values of the Zeeman term, $h=0, 0.5, 1$, is in Fig.~\ref{Fig.Conductivity40}. We already know from the density of states in the non-interacting case, Fig.~\ref{Fig.SelfDos}a, that when $h=0$, lattice effects become important only around $\omega\sim\pm 2$. This explains why the spectrum is essentially independent of $h$ for frequency less than about $\omega= 4$. Indeed, adding a Zeeman term to the Hamiltonian in Eq.~(5) of Ref.~\onlinecite{Ashby2013a} shows that, in the continuum model, the Zeeman term only shifts the Weyl nodes in the $k_z$ direction without changing the energy spectrum.

\begin{figure}
    \centering\includegraphics[width=\linewidth]{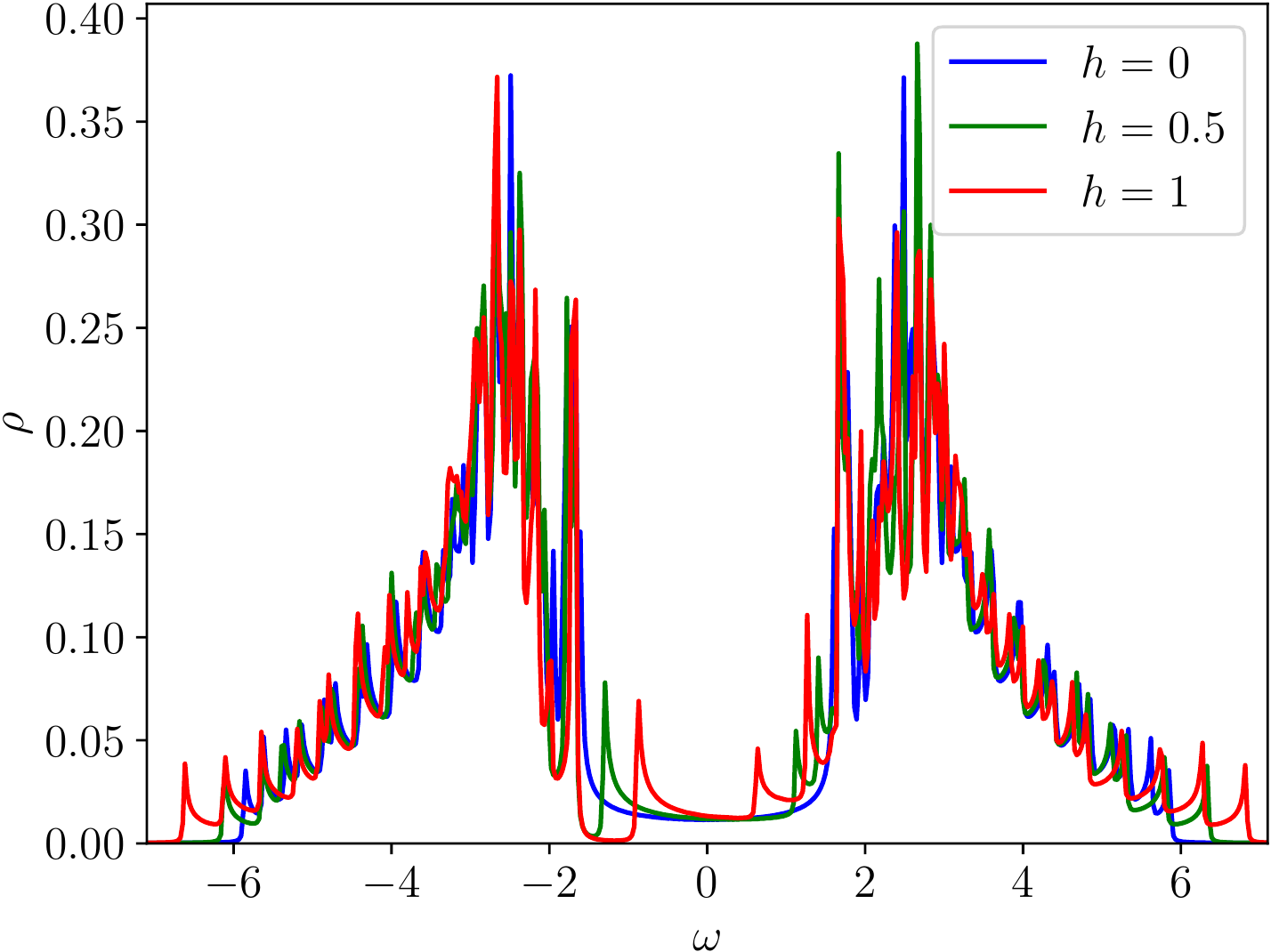}
      \caption{Local density of states for $B=2\pi/16$ and three values of the Zeeman term, $h=0, 0.5, 1$.} 
      \label{Fig.densityh}
\end{figure}

\begin{figure}
    \centering\includegraphics[width=\linewidth]{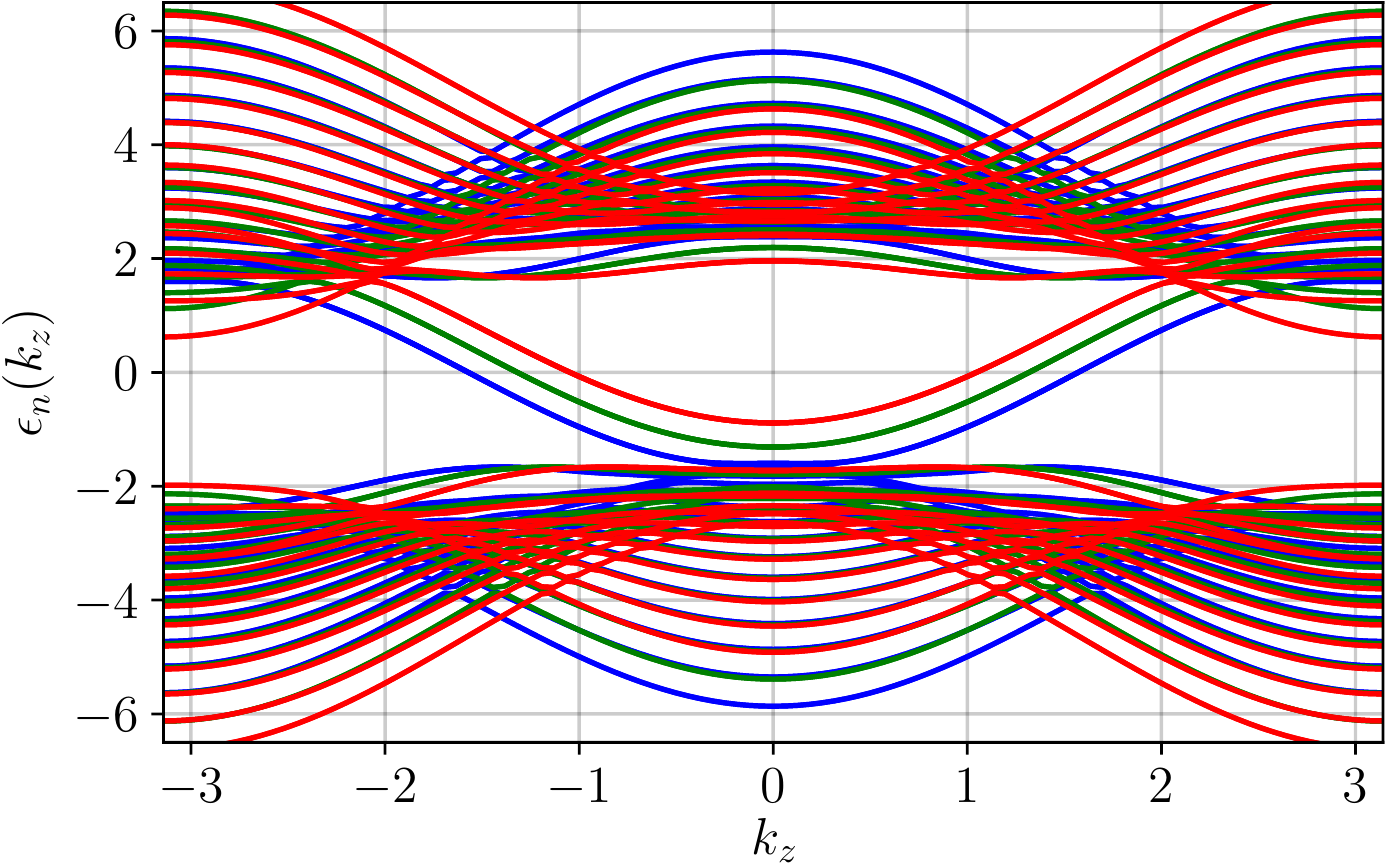}
    \caption{Landau levels for $B=2\pi/16$, and $h=0, 0.5, 1$ with the colors corresponding to the previous figure.} 
    \label{Fig.dispersionZeeman}     
\end{figure}

\begin{figure}
    \centering\includegraphics[width=\linewidth]{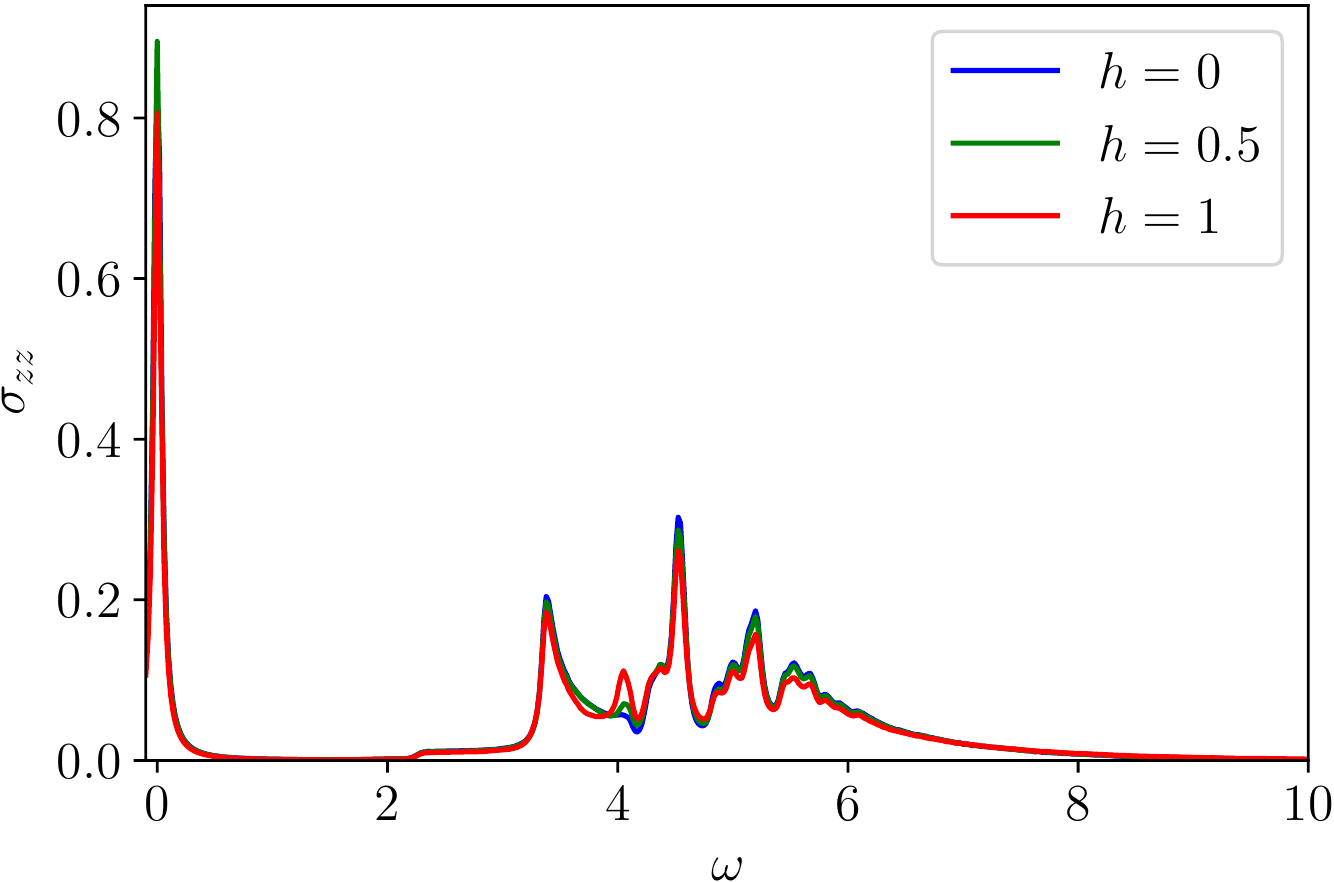}
      \caption{Magneto-optical conductivity for $B=2\pi/16$ and three values of the Zeeman term, $h=0, 0.5, 1$. Even though the Landau levels are very different for different values of $h$, the magneto-optical conductivity is not very sensitive to $h$.} 
      \label{Fig.condh}
\end{figure}

Fig.~\ref{Fig.densityh} shows the density of states for larger magnetic field, $B=2\pi/16$, and for three values of the Zeeman field $h=0, 0.5, 1$. The Landau levels are shown with the corresponding colors in Fig.~\ref{Fig.dispersionZeeman}. The difference between the three cases is striking. Note also that the magnetic field is so strong that lattice effects, as measured by the deviation from $\omega^2$ dependence of the density of states, appear at lower frequency for the largest Zeeman term. Nevertheless, the magneto-optical spectrum, shown in Fig.~\ref{Fig.condh} is not very dependent on $h$. Landau levels in the conduction and valence band vary in synchrony and selection rules enforce this near $h$ independence of the spectrum, even though lattice effects are more noticeable than for lower $B$ fields. Compared with $B=2\pi/40$, the weight of the interband transitions has decreased compared with the Drude peak and deviations from the continuum model appear at lower frequency.    

\begin{figure}
    \centering\includegraphics[width=\linewidth]{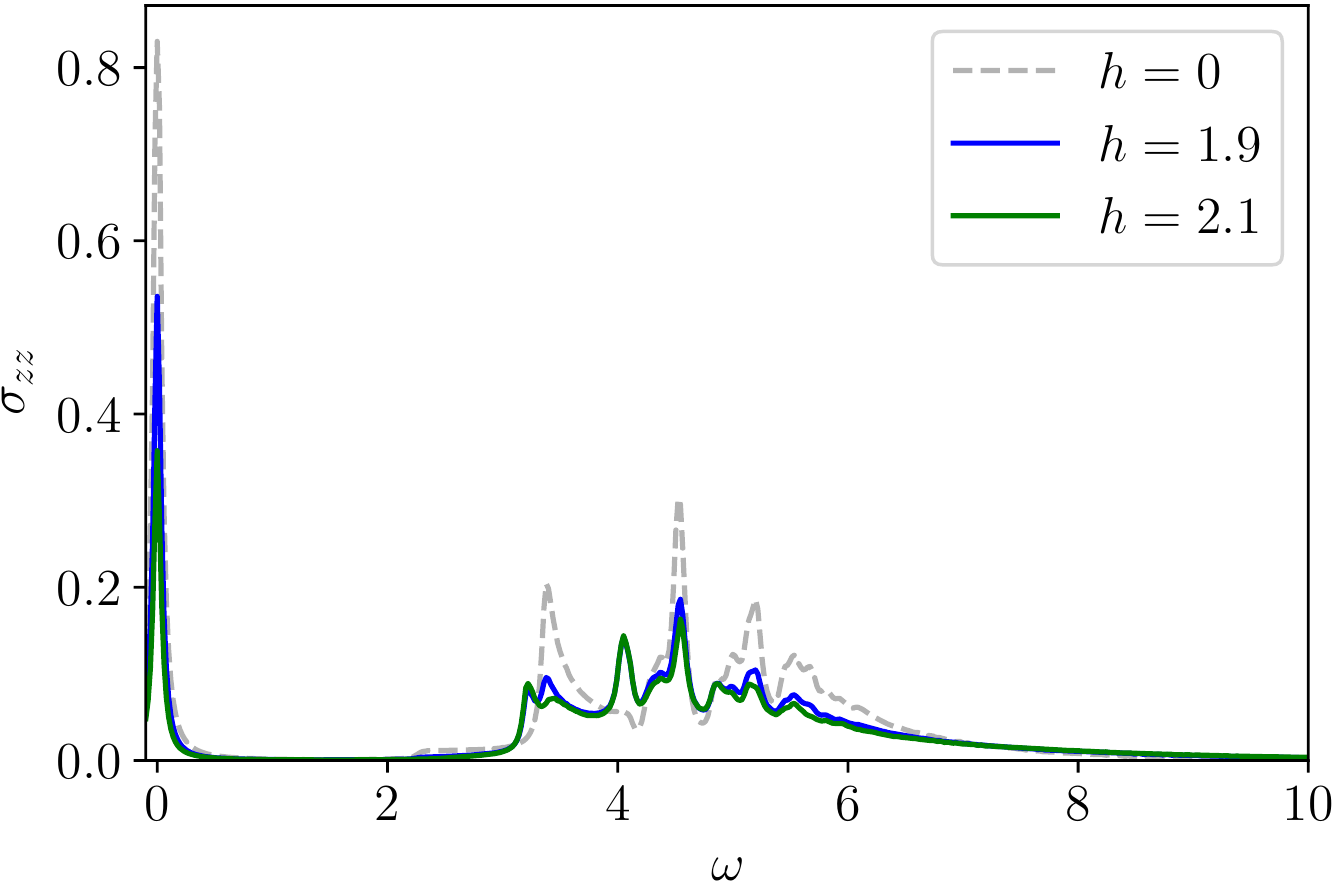}
      \caption{Magneto-optical conductivity for $B=2\pi/16$ and three values of the Zeeman term, $h=0, 1.9, 2.1$. There is a topological transition from four to two Weyl nodes at $h=2$, but nevertheless, the spectrum dos not change much across the transition.}
      \label{Fig.condtrans} 
\end{figure}

Finally, consider the extreme case near $h=2$ where there is a topological transition from four to two Weyl nodes. The results for the magneto-optical conductivity are in Fig.~\ref{Fig.condtrans}. While there is clearly a difference between $h=0$ and $h=1.9$, the change across the transition, from $h=1.9$ to $h=2.1$, is barely noticeable.   

\section{Perturbation theory of the Self-energy in the DMFT framework}\label{Appendix1}

In order to obtain Eq.~\ref{Eq.selfpert}, we use second order perturbation
theory for the self-energy. The latter is given by the formula:
\begin{equation}
\Sigma''_\omega (\omega) = -\pi U^2\int^0_{-\omega}\text{d}\epsilon_1 \int^{\omega+\epsilon_1}_0\text{d}\epsilon_2 \rho(\epsilon_1)\rho(\epsilon_2)\rho(\omega+\epsilon_1-\epsilon_2),
\end{equation}
where $\rho$ is the non-interacting local density of state of the impurity. This
is justified in the limit $U\rightarrow 0$. Using the density of states of the
low-energy Hamiltonian of a Weyl semimetal 
\begin{equation}
\rho(\epsilon) = \frac{\epsilon^2}{2\pi^2v^3_F}
\end{equation}
leads to formula~\ref{Eq.selfpert}, which shows that the imaginary part of the
self-energy scales like $\omega^8$. 

It seems that the zero-temperature Weyl semi-metal is a sort of "super Fermi
liquid" with an imaginary part that is even smaller than the $\omega^2$ of a
Fermi liquid. Nevertheless, there is a quasiparticle spectral weight that is
smaller than unity, as in Fermi-liquid theory. To show this, start from the
Kramers-Kronig relations 
imaginary parts of self-energy
\begin{equation}\label{Eq.KK}
\Sigma' (\omega) = \mathcal{P}\int \frac{\text{d} \tilde{\omega}}{\pi}\frac{\Sigma''(\tilde{\omega})}{\tilde{\omega}-\omega},
\end{equation}
where $\mathcal{P}$ stands for Cauchy's principal value. We can solve this
Eq.~\ref{Eq.KK} by using the identity
\begin{equation}
\Sigma' (\omega) = \mathcal{P}\int \frac{\text{d} \tilde{\omega}}{\pi}\frac{\Sigma''(\tilde{\omega}) - \Sigma''(\omega)}{\tilde{\omega}-\omega} + \Sigma''(\omega)\mathcal{P}\int \frac{\text{d} \tilde{\omega}}{\pi}\frac{1 }{\tilde{\omega}-\omega}.
\end{equation}

Noting that 
\begin{equation}
\tilde{\omega}^8-\omega^8=( \tilde{\omega} - \omega) ( \tilde{\omega} + \omega) ( \tilde{\omega}^2 + \omega^2) ( \tilde{\omega}^4 + \omega^4)
\end{equation}
and using a particle-hole symmetric model, we find
\begin{align}\label{Eq:ScatteringTimeReal}
\Sigma' (\omega) = &-\frac{ U^2}{40320 \pi^6 v_F^9}\bigg(\frac{2D^7 \omega}{7} +\frac{2D^5 \omega^3}{5}\nonumber \\ &+ \frac{2D^3 \omega^5}{3} +2D\omega^7 - \omega^8 \ln ( \left|\frac{\omega -D}{\omega +D} \right|)  \bigg)
\end{align}            
with $D$ the bandwidth of our toy model. Equation~\ref{Eq:ScatteringTimeReal}
has a term linear in frequency that dominates at low frequency, other terms
being of higher order in $(\omega/D)^2$. This behavior affects the quasiparticle
weight the same way as in Fermi liquid theory since
\begin{equation}
Z^{-1} = 1-\left. \frac{\partial  \Sigma'(\omega)}{\partial \omega}\right|_{\omega=0}.
\end{equation}
This proof can clearly be generalized to models without particle-hole symmetry
and for $\Sigma'' (\omega)\sim -\omega^n$ with $n$ an arbitrary integer because
$(\tilde{\omega} - \omega)$ is always a factor of $(\tilde{\omega}^n -
\omega^n)$. 

\section{Derivation of the Drude weight}\label{Appendix}

Here we show how the Drude Weight can be extracted from the imaginary-frequency
results for $\Pi_{zz}(i\nu_n)$. Taking advantage of the fact that there is a gap
in the optical conductivity, we can write it as a sum of Drude and of other
contributions:
\begin{equation}
\sigma'_{zz} = \sigma^D_{zz} + \sigma^{res}_{zz}.
\end{equation}
Using gauge invariance, the f-sum rule~\cite{Maldague_1977} can be written as
follows
\begin{align}\label{Eq:fsum}
\int \text{d}\omega \sigma'_{zz}(\omega) &=\int \text{d}\omega \sigma^D_{zz}(\omega) + \int \text{d}\omega \sigma^{res}_{zz}(\omega)\nonumber \\&= \int \text{d}\omega \frac{\Pi''_{zz}(\omega)}{\omega} = \pi \Pi_{zz}^{para}(i\nu_n=0) ,
\end{align}
where $\Pi_{zz}^{para}$ is the paramagnetic contribution. 

In Weyl semimetals subject to an external magnetic field, a selection rule on
the conductivity along the direction of the magnetic field
applies, as can be deduuced from Ref.~\onlinecite{Ashby2013a}. Indeed, with $n$ being the index of the
$n^{\text{th}}$ Landau, the only possible transitions are $-|n|\rightarrow |n|$.
Then the contribution to the conductivity of the zeroth Landau level appears
only very near zero frequency. This selection rule allows us to do a gedanken
experiment and imagine that the zeroth Landau level is not present anymore. Then
the f-sum rule becomes
\begin{equation}
\int \text{d}\omega \sigma^{res}(\omega) = \pi \tilde{\Pi}^{para}_{zz}(i\nu_n=0),
\end{equation}
with $\tilde{\Pi}^{para}_{zz}(i\nu_n=0)$ the \textit{paramagnetic part of the
correlation function in the absence of the Drude contribution}. In the absence
of this contribution, there is a gap in the optical conductivity, so the
spectral representation
\begin{eqnarray}
    \tilde{\Pi}^{para}_{zz}(i\nu_n)=\int \frac{\text{d}\omega}{\pi} \frac{\Pi_{zz}^{\prime\prime \ no  \  Drude}(\omega)}{\omega-i\nu_n}
\end{eqnarray}
tells us that at temperatures smaller than the gap,
$\tilde{\Pi}^{para}_{zz}(i\nu_n)$ can be extrapolated to $\nu_n=0$ from a
polynomial fit of $\Pi^{Para}_{zz}(i\nu_n \neq 0)$. Once
$\tilde{\Pi}^{Para}_{zz}$ is computed, the Drude weight is easily obtained using
Eq.~\ref{Eq:fsum}
\begin{align}\label{Eq.DrudeWeightEq}
\int \text{d}\omega &\sigma^D_{zz}(\omega) = \pi \left(\Pi^{Para}_{zz}(i\nu_n = 0) -  \tilde{\Pi}^{Para}_{zz}(i\nu_n = 0)\right).
\end{align}
Given the remarkable robustness of quasiparticle physics in Weyl semimetal at
low temperature, we can use this result even for large values of $U$, as long as
there is a gap between the Drude peak and the interband transitions.



%

\end{document}